\documentclass[10pt,conference]{IEEEtran}
\IEEEoverridecommandlockouts
\usepackage{cite}
\usepackage{amsmath,amssymb,amsfonts}
\usepackage{algorithmic}
\usepackage{graphicx}
\usepackage{textcomp}
\usepackage{xcolor}
\usepackage{subcaption}
\usepackage{graphicx}
\usepackage{xcolor}
\usepackage{braket}
\usepackage{dcolumn}
\usepackage{amsmath} 
\usepackage{bm}
\def\BibTeX{{\rm B\kern-.05em{\sc i\kern-.025em b}\kern-.08em
    T\kern-.1667em\lower.7ex\hbox{E}\kern-.125emX}}
\begin{document}

\title{\fontsize{17}{15}\selectfont Long-Range Entangling Operations via Josephson Junction Metasurfaces
\thanks{M. Bakr acknowledges support from EPSRC QT Fellowship Grant EP/W027992/1.}
}
\author{\IEEEauthorblockN{Mustafa Bakr\textsuperscript}
\IEEEauthorblockA{\textit{Clarendon Laboratory, Department of Physics}\\
\textit{University of Oxford}
Oxford, UK \\
mustafa.bakr@physics.ox.ac.uk}
}

\maketitle

\begin{abstract}
We present a framework for implementing two-qubit entangling operations between distant superconducting qubits using a space–time modulated Josephson junction (JJ) metasurface. By modulating the surface in both space and time, we engineer sidebands with controllable wavevectors that selectively couple target qubits. The metasurface acts as a reconfigurable coupling medium, where the interaction strength is determined by engineered transmission coefficients \( T_\mu(\mathbf{k}_\mu \cdot \mathbf{r}) \) rather than by exponentially decaying near-field coupling, thus reducing the dependence on physical proximity. We investigated the implementation of two-qubit interactions via iSWAP gates driven resonantly through the metasurface and controlled phase gates via geometric phase accumulation. Simulations show entangling fidelity exceeding 98\% maintained over centimeter-scale separations.

\end{abstract}


\section{Introduction}
High-fidelity two-qubit entangling operations are essential for scalable quantum computing, yet conventional architectures face fundamental trade-offs between connectivity, speed, and fidelity. The Mølmer--Sørensen (MS) gate \cite{Sorensen2000} enables high-fidelity entanglement via off-resonant, phase-controlled sidebands coupling to a shared bosonic mode. We generalize this concept using space–time modulated JJ metasurfaces \cite{Bakr2025}, allowing sideband excitation over spatially distributed electromagnetic eigenmodes with engineered wavevector control. Unlike cavity-bound implementations, our metasurface provides active beam steering, multi-qubit connectivity, and low-loss frequency multiplexing. Crucially the interaction strength is mediated through engineered transmission coefficients rather than exponentially decaying near-field coupling, fundamentally altering the distance dependence. This could potentially enable all-to-all connectivity within a chip or across modules, supporting fault-tolerant quantum computing with low-density parity-check (LDPC) codes. These codes demand rapid, parallel nonlocal stabilizer measurements that are infeasible in nearest-neighbour architectures. By using metasurface-mediated gates to bridge distant qubits without increasing thermal or wiring load, our architecture provides a potential route to scalable, error-corrected superconducting quantum processors.

\section{Theoretical Framework}
We consider a system of \( N \) fixed-frequency superconducting qubits located at positions \( {r}_i \), each coupled to a spatially extended, time-modulated JJ metasurface described in Fig. 1 \& 3 of [2], which supports quantized electromagnetic modes \( b_\mu \) with dispersion relation \( \nu_\mu \) and is engineered to provide controllable coupling between spatially separated qubits through Floquet sideband generation. The time-dependent Hamiltonian in the interaction picture is given by
\begin{equation}
H_I(t) = \sum_\mu \epsilon_\mu b_\mu^\dagger b_\mu + \sum_{i,\mu} g_{i\mu}(t) \left( \sigma_i^+ b_\mu e^{i(\omega_i - \nu_\mu)t} + \text{h.c.} \right),
\label{eq:hamiltonian_interaction}
\end{equation}
where \( \epsilon_\mu = \nu_\mu - \omega_d \) denotes the mode detuning from the reference drive.

The coupling coefficients are position-dependent and take the form
\begin{equation}
g_{i\mu}(t) = g_i^{(0)} T_\mu({k}_\mu \cdot {r}_i) e^{i {k}_\mu \cdot {r}_i},
\label{eq:coupling}
\end{equation}
where \( g_i^{(0)} \) is the bare coupling strength, \( T_\mu \) is the transmission amplitude encoding beam propagation, focusing, and spatial selectivity, and the exponential phase term encodes spatial coherence.

To simplify the dynamics, we apply local qubit rotations defined by \( U_i = \exp(-i \theta_i Y_i / 2) \) with \( \tan \theta_i = \Omega_i / \Delta_i \), where \( \Omega_i \) and \( \Delta_i \) represent qubit drive amplitude and detuning, respectively. In the dressed qubit basis~\cite{Gorshkov2025}, the effective Hamiltonian becomes
\begin{equation}
H_{\text{eff}} = \sum_\mu \epsilon_\mu b_\mu^\dagger b_\mu + \sum_{i,\mu} \frac{g_{i\mu}\sin \theta_i}{2} Z_i (b_\mu + b_\mu^\dagger).
\label{eq:hamiltonian_effective}
\end{equation}
In this frame, each qubit acts as an effective spin in a state-
dependent linear potential, enabling geometric phase accumulation through closed trajectories in the joint qubit-mode phase space. We analyze two distinct interaction schemes enabled by metasurface modulation.

\subsection{iSWAP interaction via metasurface drives}
We modulate the metasurface to create effective exchange coupling between qubits through parametric modulation of the JJ array. The space–time modulated metasurface enables direct \( XX + YY \) interactions without requiring separate microwave sources. Under the rotating wave approximation and assuming dominant mode contributions, the effective two-qubit Hamiltonian becomes
\begin{equation}
H_{\text{iSWAP}} = \frac{J_{\text{eff}}}{2}(X_1X_2 + Y_1Y_2) - \frac{\Delta}{2}(Z_1 - Z_2),
\label{eq:iSWAP}
\end{equation}
where \( \Delta = \omega_1 - \omega_2 \) is the inter-qubit detuning. The effective exchange coupling strength is determined by the metasurface-mediated interaction
\begin{equation}
J_{\text{eff}} = \frac{g_1^{(0)} g_2^{(0)}}{2} \left| T_\mu(k_\mu \cdot r_1) \right| \left| T_\mu(k_\mu \cdot r_2) \right| \cos(k_m(z_1 - z_2)),
\label{eq:J_eff}
\end{equation}
where \( g_i^{(0)} \) are the bare qubit–metasurface coupling strengths, \( T_\mu(k_\mu \cdot r_i) \) are the engineered transmission coefficients encoding spatial selectivity, and \( \cos(k_m(z_1 - z_2)) \) captures the modulation-induced phase relationship between qubits.
\begin{figure}[!t]
	\begin{center}
	\includegraphics[width=0.9\columnwidth]{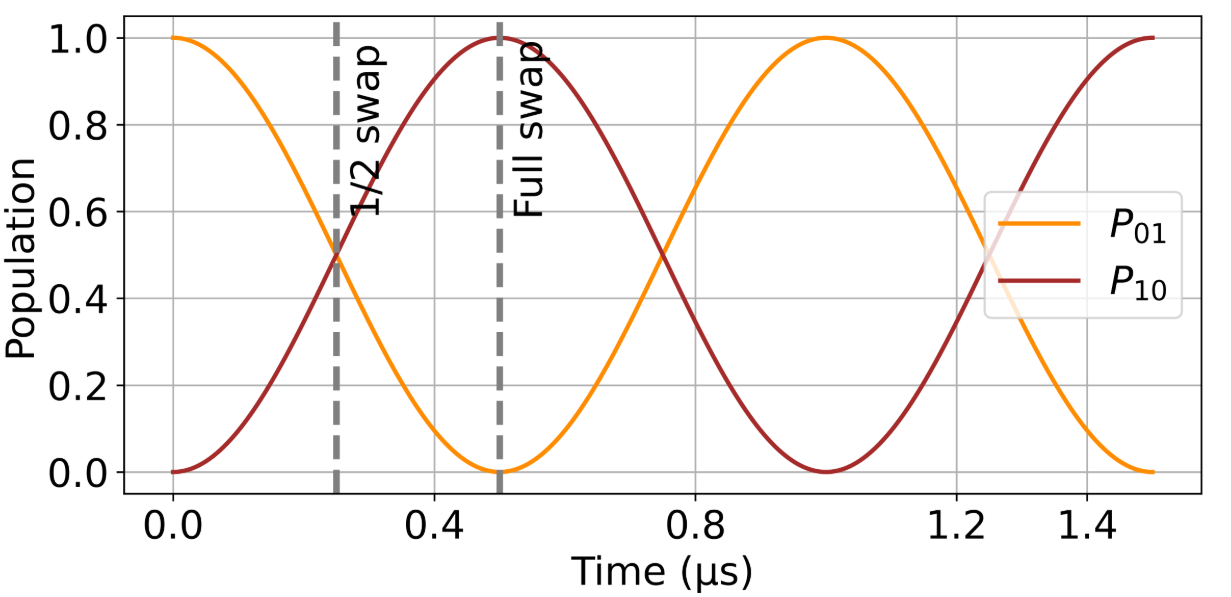}
		\caption{Exchange interaction dynamics under JJ metasurface modulation, demonstrating coherent oscillation of excitation between states \( \lvert 01 \rangle \) and \( \lvert 10 \rangle \).}
		\label{figcc}
	\end{center}
\end{figure}
The metasurface modulation creates time-dependent coupling coefficients:
\begin{equation}
g_{i\mu}(t) = g_i^{(0)} T_\mu(k_\mu \cdot r_i) e^{i k_\mu \cdot r_i} \delta(t),
\label{eq:coupling_modulation}
\end{equation}
where \( \delta(t) \) represents the parametric modulation envelope of the JJ array. When the exchange coupling dominates (\( J_{\text{eff}} \gg |\Delta| \)), the interaction reduces to the pure iSWAP form 
\begin{equation}
H_{\text{eff}} \approx \frac{J_{\text{eff}}}{2}(X_1X_2 + Y_1Y_2).
\label{eq:iswap_final}
\end{equation}
This mediates coherent population exchange between states \(\ket{01}\leftrightarrow\ket{10}\), as illustrated in Fig.~1, validating the proposed metasurface-induced interactions.

\subsection{Controlled-phase via geometric phases}
Controlled-phase interactions are realized via geometric phases accumulated in the dressed qubit basis. The interaction duration \( \tau \) required for a conditional \( \pi \)-phase shift between qubits \( i,j \) is determined by

\begin{equation}
\sum_\mu \frac{(g_{i\mu}\sin\theta_i + g_{j\mu}\sin\theta_j)^2}{4\epsilon_\mu}\tau = \pi.
\label{eq:gate_condition}
\end{equation}
The effective long-range inter-
action strength is determined by
\begin{equation}
|J_{ij}| = \sum_\mu \frac{|g_i^{(0)} g_j^{(0)}T_\mu({k}_\mu \cdot {r}_i)T_\mu({k}_\mu \cdot {r}_j)|}{4|\epsilon_\mu|},
\label{eq:J_eff_cp}
\end{equation}
The transformation introduces additional dispersive terms
\begin{equation}
H_{\text{disp}} = \sum_{i,\mu} f_{i\mu}\, Z_i\, b_\mu^\dagger b_\mu, \quad \text{with } f_{i\mu} = \frac{|g_{i\mu}|^2}{\Delta_i}, \label{eq:dispersive}
\end{equation}
where the dispersive coupling strength $f_{i,\mu} = |g_{i,\mu}|^2/\Delta_i$ now includes the spatial transmission factor. This creates position-dependent dispersive shifts that must be carefully managed to maintain gate fidelity. These frequency shifts can compromise interaction fidelity, necessitating dynamical decoupling protocols. Optimal error suppression occurs when \(\epsilon_\mu\tau=2\pi n\), achievable through precise modulation control. The interaction time scales as
\begin{equation}
t_{\text{interaction}} \sim \frac{\pi}{4 |J_{ij}|} \sim \frac{\pi |\epsilon_\mu|}{|g_i^{(0)} g_j^{(0)}|\, |T_\mu|^2}, \label{eq:gate_time}
\end{equation}
Note that the coupling strength depends critically on the transmission coefficients \( T_\mu \), which can be engineered through metasurface design to maximize \( |T_\mu(\mathbf{r}_i)||T_\mu(\mathbf{r}_j)| \) for target qubits while minimizing coupling elsewhere.
\begin{figure}[!t]
	\begin{center}
	\includegraphics[width=1\columnwidth]{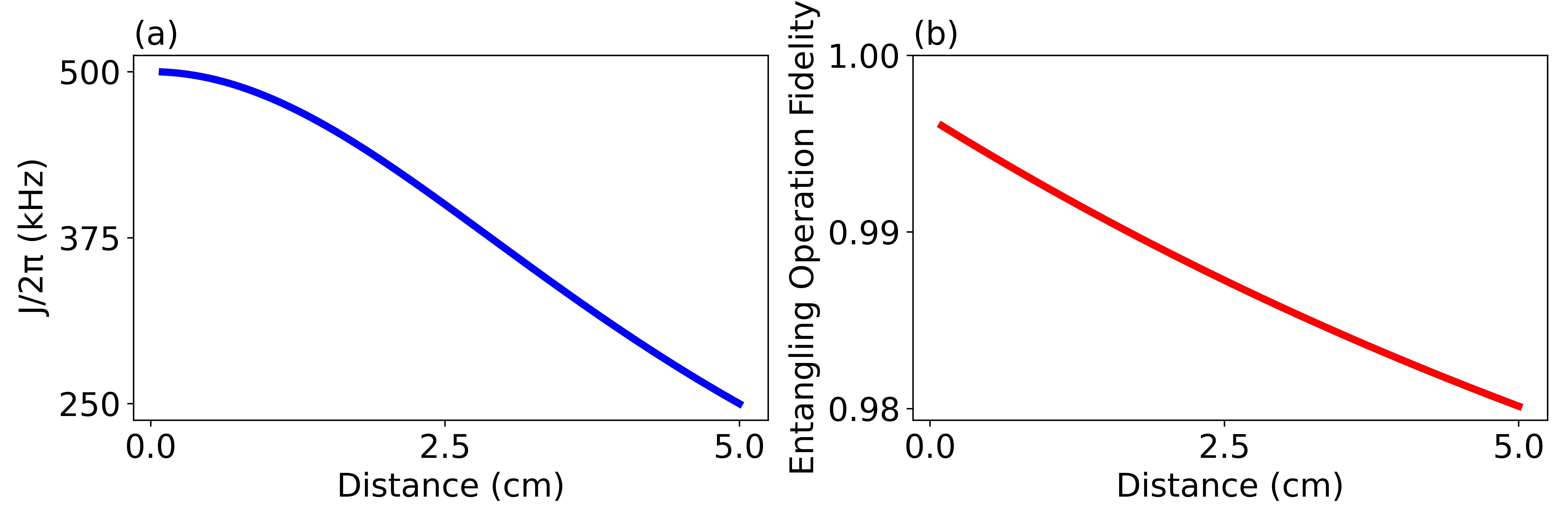}
		\caption{(a) Metasurface-mediated coupling strength as a function of qubit separation. (b) Entangling operation, maintaining values above 98\% across separations up to 5 cm. }
		\label{figcc}
	\end{center}
\end{figure}
Numerical results demonstrate weak distance scaling, with interaction strengths reducing by only a factor of two over distances of several centimeters. Realistic parameters (\(g_i^{(0)}/2\pi = 28.3\,\text{MHz}\), \(|\epsilon_\mu|/2\pi = 200\,\text{MHz}\), \(T_1 = 500\,\mu\text{s}\), \(t_{\text{gate}} \approx 250\,\text{ns}\)) yield fidelities exceeding 98\%, assuming metasurface transmission efficiency \(|T_\mu|^2>0.5\) and mode isolation above 23\,dB (Fig.~2). This highlights a fundamental improvement over exponential near-field coupling limitations, enabled by engineered wavevector control via metasurface modulation.

\section{Conclusion}
We have proposed and analyze dynamically modulated JJ metasurfaces to mediate long-range two-qubit entangling operations.  By leveraging spatially engineered electromagnetic mode coupling, we reduce the limitations imposed by traditional near-field exponential coupling decay. Our simulations indicate robust fidelities exceeding 98\% over centimeter-scale distances. This approach promises a potential practical route toward implementing quantum error correction protocols, such as LDPC codes. 

\bibliographystyle{IEEEtran}

\end{document}